\documentclass[aps,prb,twocolumn,showpacs]{revtex4}

\usepackage{amsmath,amssymb,dsfont,citesort,graphicx,psfrag}

\newcommand{\crs}{CeRu$_2$Si$_2$}
\begin{document}

\title{Universal signatures of the metamagnetic quantum critical endpoint: Application to CeRu$_2$Si$_2$}

\author{Franziska Weickert}
\author{Manuel Brando}
\author{Frank Steglich}
\affiliation{Max-Planck-Institut f\"ur Chemische Physik fester
Stoffe, 01187 Dresden, Germany}

\author{Philipp Gegenwart}
\affiliation{I. Physikalisches Institut, Georg-August-Universit\"at
G\"ottingen, 37077 G\"ottingen, Germany}

\author{Markus Garst}
\affiliation{Institut f\"ur Theoretische Physik, Universit\"at zu K\"oln, 50938 K\"oln, Germany\\
and Physik Department, Technische Universit\"at M\"unchen, 85748
Garching, Germany. }

\date{\today}
\begin{abstract}
A quantum critical endpoint related to a metamagnetic transition
causes distinct signatures in the thermodynamic quantities of a
compound. We argue that, irrespective of the microscopic details of
the considered material, the diverging differential susceptibility
combined with the Ising symmetry of the endpoint give rise to a
number of characteristic metamagnetic phenomena.
In the presence of a magnetoelastic coupling, one finds a correspondence of susceptibility, magnetostriction and compressibility and, as a result, a pronounced crystal softening, a diverging Gr\"uneisen parameter, a sign change of thermal expansion $\alpha(H)$, and a minimum in the specific heat coefficient $\gamma(H)$. We illustrate these signatures and their relation on the
metamagnetic crossover at 8\,T in the prototypical heavy-fermion
system CeRu$_2$Si$_2$.
\end{abstract}
\pacs{75.30.Kz; 71.27.+a; 71.10.HF}
\maketitle
\section{Introduction}
Emergent universality close to critical points is one of the most
fascinating phenomena in physics. Materials with utterly distinct
microscopic composition might exhibit similar behavior close to a
second order phase transition if they only belong to the same
universality class. Universality is also expected close to a
critical endpoint that terminates a line of first-order transitions
like, for example, in the phase diagram of the liquid-gas
transition. Such an endpoint is characterized by Ising universality,
though the Ising order parameter is sometimes not simply related to
measureable quantities.

Another example of an Ising critical endpoint can be found in
metamagnetic materials. Metamagnetism is often casually described as
a superlinear rise of the magnetization $M(H)$ at some finite
critical field $H_m$. With decreasing temperature, such a smooth
metamagnetic crossover might evolve into a sharp first-order jump of
$M(H)$ identifying an endpoint $(T_{\rm ep},H_m)$ in the phase
diagram plane defined by temperature, $T$, and magnetic field
$H$.\cite{Wolfarth62,Yamada93} In the context of the metamagnetic
material Sr$_3$Ru$_2$O$_7$ it was pointed
out\cite{Grigera01,Millis02} that an interesting situation arises if
the endpoint temperature $T_{\rm ep}$ can be tuned towards zero,
$T_{\rm ep} \to 0$, by a certain external control parameter resulting in a
quantum critical endpoint (QCEP) for $T_{\rm ep}=0$. Such an Ising QCEP differs from its classical counterpart at finite
$T_{\rm ep}$ because the dynamics of the Ising order parameter has
to be taken into account explicitly in the quantum
case.\cite{Millis02} This dynamics then generate the temperature
dependence in its vicinity giving rise to quantum critical scaling.
For example, the differential susceptibility at $H_m$ diverges by
definition with decreasing temperature with a characteristic
powerlaw, $\chi \sim T^{-x}$. Despite the fact that a QCEP can be
hardly realized in any material due to the required fine-tuning
$T_{\rm ep} = 0$, it might nevertheless control thermodynamics in an
extended temperature and field range if it is only close in
parameter space. This motivates us to look for universal signatures
of the metamagnetic QCEP in various metamagnetic compounds.

In the present work, we concentrate on the canonical heavy-fermion
material CeRu$_2$Si$_2$, which crystallizes in the tetragonal
ThCr$_{2}$Si$_{2}$-structure and shows a pronounced metamagnetic
crossover for field parallel to the crystallographic $c$-direction
with a steep rise in the magnetization $M(H)$ at $\mu_{0}H_m \approx
8\,$T.\cite{Haen87} Over the last 20 years, its properties close to
the critical field $H_m$ have been intensively investigated by
various experimental methods promoting it to be one of the
best-studied metamagnetic metals, cf.~the review
Ref.~\onlinecite{Flouquet05}. It has been noted early on that the
metamagnetic signatures are mirrored by strong anomalies in
dilatometry due to magnetoelastic coupling.\cite{Flouquet95} Close
to $H_m$ one observes a remarkable large Gr\"uneisen
parameter,\cite{Lacerda89} a sign change of the thermal
expansion\cite{Lacerda89b} and a strong crystal
softening.\cite{Kouroudis87} Moreover, the specific heat coefficient
$\gamma(H)$ shows a characteristic double peak structure close to
the critical field $H_m$.\cite{Aoki98} Although the temperature
dependence of thermodynamics is anomalous, Fermi liquid behavior is
recovered at lowest temperatures for all magnetic fields. In
particular, the differential susceptibility at $H_m$ first increases
with decreasing $T$ but then starts to saturate at a temperature of
the order of $T^* = 0.5\,$K.\cite{Holtmeier95,Sakakibara95}
In the past, the experimental results have been often interpreted within a scaling-ansatz for the entropy of the form $S(H/H_m(p), T/T_0(p))$, with the magnetic field $H$, temperature $T$, pressure dependent critical field $H_m(p)$ and the temperature scale $T_0(p)$.\cite{Lacerda89,Paulsen90,Matsuhira99,Matsuhira99b} This phenomenological approach was quite successful to account for the observed relations between the $H$-dependence of various thermodynamic quantities in the low-temperature limit. However, it did not provide an explanation for the huge anomalies themselves like, for example, the large Gr\"uneisen parameter. Since then, a number of microscopic theories based on the periodic Anderson or Hubbard model have been put forward for metamagnetism in heavy-fermion materials\cite{Ohkawa89,Satoh01,Edwards97,Meyer01} that qualitatively explained many of the observed features in CeRu$_2$Si$_2$.

Nevertheless, in a previous comparison\cite{Flouquet02} of CeRu$_2$Si$_2$ with the ruthenate Sr$_3$Ru$_2$O$_7$ it was already noted that both compounds share similar metamagnetic anomalies despite their microscopic differences. This encourages us to speculate that some of these anomalies are not specific to the microscopics of CeRu$_2$Si$_2$ but are, in fact,  associated with the emergent universality expected close to a metamagnetic QCEP. Note that the importance of critical magnetic fluctuations in this material has been anticipated early on.\cite{Lacerda89b,Paulsen90} In the following, we demonstrate that, actually, some of the most striking metamagnetic features in CeRu$_2$Si$_2$ can be naturally explained within a QCEP scenario.
We argue that the emergent Ising symmetry combined with the enhanced differential susceptibility give rise to the following generic phenomena close to a metamagnetic QCEP: (1) a correspondence between susceptibility, magnetostriction and elastic constants, and, as a consequence, (2) a pronounced crystal softening, (3) an enhanced Gr\"uneisen parameter, (4) a sign change of the thermal expansion, and (5) a minimum of the specific heat coefficient, $\gamma(H)$, with two accompanying side peaks.
Note, however, that the QCEP itself is not realized, neither in
Sr$_3$Ru$_2$O$_7$ nor in CeRu$_2$Si$_2$. Whereas in the ruthenate it
is masked by a thermodynamic phase,\cite{Grigera04} the saturation
of the differential susceptibility in CeRu$_2$Si$_2$ and the
concomitant onset of Fermi liquid behavior indicates that the QCEP
is close but still off in parameter space, its distance measured by
the saturation temperature $T^*$. For this reason the sharp increase
in the magnetization in \crs\ is mostly called
\textit{metamagnetic-like} transition or \textit{metamagnetic
crossover} in the literature.\cite{Haen87} It is an open question
whether there exists any tuning parameter that lowers the saturation temperature $T^*$ further to bring \crs\ closer to quantum criticality. Attempts to tune the
system to the QCEP or even to a first order phase transition by
applying pressure\cite{mignot88} or Ge-doping on the
Si-site,\cite{Weickert05} the latter being equivalent to negative
pressure, failed. In both cases the thermodynamic signatures at the
metamagnetic crossover are even broader compared to the pure system at ambient pressure.

We would like to point out that the concept of a metamagnetic QCEP should be distinguished from a quantum critical point (QCP) that separates two phases with different symmetry. The QCEP is the endpoint of a line of first-order transitions that is tuned to zero temperature, and as such it is an isolated singular point in the phase diagram that is surrounded by a single thermodynamic phase. 
In contrast, a QCP separates different thermodynamics phases, for example, a paramagnet and an antiferromagnet. The latter is actually realized in the Ge-doped sister compound CeRu$_2($Si$_{1-x}$Ge$_x)_2$ close to concentrations $x \simeq 0.06-0.07$.\cite{Haen99}


For the comparison presented here, we performed new measurements on
a single crystal of CeRu$_{2}$Si$_{2}$ grown by Czochralski
method.\cite{Grenoble} Thermal expansion, magnetostriction and heat
capacity measurements were carried out in a $^{3}$He/$^{4}$He
dilution refrigerator mounted inside an 18~T superconducting magnet.
A capacitive dilatometer\cite{Pott83} made of CuBe and with high
sensitivity of $\frac{\Delta L}{L_{0}} = 10^{-10}$ was used to
estimate the thermal expansion coefficient $\alpha_{c}(H,T) =
\frac{1}{L_{c0}}\frac{\partial L_{c}}{\partial T}$ and the
magnetostriction coefficient $\lambda_{c}(H,T) =
\frac{1}{L_{c0}\mu_{0}}\frac{\partial L_{c}}{\partial H}$ for $L
\parallel c$. It is well known that in \crs\ the length change of
the sample $\Delta L$ shows the same temperature and field
dependence along all crystallographic axes, but $\Delta L_c$ is
three times larger than  $\Delta L_a$.\cite{Paulsen90}
For this reason, we scaled our data by a factor of $\frac{5}{3}$ to
get the volume expansion $\alpha$ and magnetostriction $\lambda$.
The heat capacity was measured by a silver platform using
compensated heat-pulse technique.\cite{Wilhelm04} In both
experiments we tuned the magnetic field very close to the
metamagnetic critical field $H_m$ in ultrafine field steps of just a
few mT. We were able to extend the temperature range of our
measurements down to 60\,mK in comparison to former publications.
Our results reproduce nicely existing data for the same temperatures
and fields of $\alpha$,
$\lambda$\cite{Paulsen90,Lacerda89,Matsuhira99,Matsuhira99b} and the
specific heat.\cite{Fisher88,Aoki98,Heuser00}

The paper is organized as follows. In Section \ref{sec:unisig} we
discuss qualitatively the characteristic thermodynamics expected
close to a metamagnetic QCEP, in Section \ref{sec:comparison} we
discuss the metamagnetic signatures and their relationship in
CeRu$_2$Si$_2$, and we end with a summary in section
\ref{sec:summary}.

\section{Universal signatures close to a metamagnetic QCEP}
\label{sec:unisig}

\begin{figure}
\includegraphics[width=0.35\textwidth]{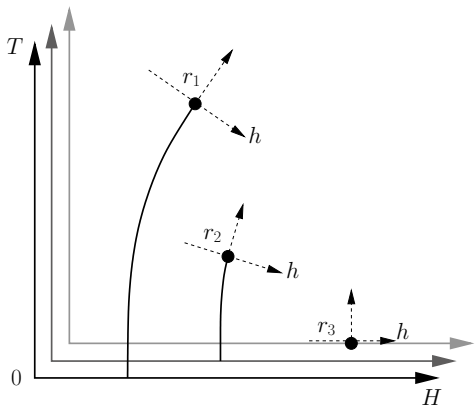}
\caption{Schematic depiction of planes in the $H-T$-phase diagram with a critical endpoint, which terminates a line of a metamagnetic
first-order transition. The parameter $r$ is a measure of the
distance to the QCEP in parameter space, and three different cases
$r_{i}$, $r_{1}< 0$, $r_{1}<r_{2}<0$, $r_{3}=0$ are shown. For $r>0$ the endpoint disappears below the $T=0$ axis, which corresponds to the situation in \crs. The
dashed local coordinate system close to the endpoint aligns itself
with the physical coordinates as the endpoint temperature approaches
zero, $T_{\rm ep} \to 0$. } \label{fig:endpoint}
\end{figure}

Close to a metamagnetic quantum critical endpoint the free energy density, $\mathcal{F} = \mathcal{F}_0 + \mathcal{F}_{\rm cr}$, can be separated into a background part, $\mathcal{F}_0$, and a critical part, $\mathcal{F}_{\rm cr}$, deriving from the Ising QCEP
\begin{align} \label{Scaling}
\mathcal{F}_{\rm cr} = \mathcal{F}_{\rm cr}(h,T, r).
\end{align}
We assume that $\mathcal{F}_0$ yields only a small, featureless
background contribution to thermodynamics that is sub-leading
compared to the critical one. The critical part (\ref{Scaling})
depends on temperature, $T$, on the magnetic scaling field $h$ that
is conjugate to the Ising order parameter, and a parameter, $r$,
that measures the distance to quantum criticality. A negative value,
$r<0$, corresponds to a finite endpoint temperature $T_{\rm ep} >0$,
and for $r>0$ only a metamagnetic crossover occurs. The QCEP is
realized exactly for $r = 0$. Additionally, it is important how the
magnetic scaling field, $h$, is related to the physical fields.
Usually for endpoints, this relation is not evident as, e.g., for
the liquid-gas transition, and, as a consequence, the Ising symmetry
is often hidden. However, in our case the situation is more
fortunate as we are dealing with an endpoint at $T=0$. In the limit
of a vanishing endpoint temperature $T_{\rm ep} \to 0$, the line of
first-order metamagnetic transitions will align itself with the
temperature axis in the $(H,T)$ phase diagram. This follows from the
Clausius-Clapeyron relation after taking into account that the
transition at $T=0$ is between two ground states of same entropy,
see Fig.~\ref{fig:endpoint}. This alignment has the consequence that
the magnetic scaling field, $h$, of the Ising QCEP can be directly
identified with the distance to the critical magnetic field
\begin{align} \label{scalingField}
\left.h\right|_{T\to 0} = H-H_m.
\end{align}
The magnetic scaling field, $h$, thus controls the distance to the
QCEP directly on the magnetic field axis and, as a result, the Ising
symmetry becomes explicit in the $(H,T)$ phase diagram. At a finite
temperature, there will be a superlinear $T$-correction to $h$ {\it
even} for the QCEP, $r=0$; we will later see that for
CeRu$_{2}$Si$_{2}$ this correction is of order $\mathcal{O}(T^2)$.

Within certain models,\cite{Yamada93,Millis02,Binz04} the critical free energy (\ref{Scaling}) can be calculated and the dependences on its parameters can be determined. In the present work, we will follow a different route and concentrate on the generic properties that all these models for a metamagnetic QCEP have in common. We will therefore focus on a qualitative discussion of the universal metamagnetic signatures that derive from two basic assumptions: (i)  a diverging differential susceptibility $\chi$ at the QCEP and (ii) its Ising universality.

Probably the most fundamental quantity, that characterizes the metamagnetic behavior, is the differential susceptibility $\chi$,
\begin{align}
\chi = - \frac{\partial^2 \mathcal{F}}{\partial H^2}.
\end{align}
At a QCEP, $r=0$, the susceptibility diverges upon decreasing temperature, $T$, or decreasing $h$,
\begin{align} \label{defQCEP}
\left. \chi \right|_{\rm QCEP}
\to \infty
\quad \mbox{as} \quad T, |h| \to 0.
\end{align}
This serves as our definition of quantum critical metamagnetism. We
show below that this increase of $\chi$ close to the QCEP is
responsible for all of the striking metamagnetic phenomena. The
pronounced increase of the susceptibility $\chi$ in
CeRu$_{2}$Si$_{2}$ at $H_m$ only saturates at a temperature $T^* =
0.5\,$K. In the QCEP scenario, this saturation temperature $T^*$ is
a measure of a finite positive parameter $r$ in Eq.~(\ref{Scaling}),
$\lim_{T,|h|\to 0} \chi \sim 1/r$.


Apart from $\chi$, we will discuss various other second order derivatives of the free energy: the specific heat coefficient $\gamma$, thermal expansion $\alpha$, magnetostriction $\lambda$ and compressibility $\kappa$, defined as
\begin{gather} \label{ThermQu}
\gamma = - \frac{\partial^2 \mathcal{F}}{\partial T^2},\quad
\alpha = \frac{\partial^2 \mathcal{F}}{\partial T \partial p},\quad
\lambda  = \frac{\partial^2 \mathcal{F}}{\partial H \partial p},\quad
\kappa = - \frac{\partial^2 \mathcal{F}}{\partial p^2}.
\end{gather}
The pressure dependence of the critical part $\mathcal{F}_{\rm cr}$
enters via the smooth pressure dependence of $h$, i.e., the critical
field, $H_m = H_m(p)$. In the language of field theory, the pressure
dependence of the other parameter $r$ is less relevant and will be
neglected in the following. It is convenient to define
\begin{align} \label{OmegaDef}
\Omega_m = \frac{\partial H_m}{\partial p}.
\end{align}
%
For the small pressures applied in dilatometric experiments
$\Omega_m$ can be approximated to be constant, $\Omega_m \approx$
const.

Because of the scaling form in Eq.~(\ref{Scaling}), the critical
contributions to thermodynamic quantities are not independent. It
follows from Eq.~(\ref{Scaling}) that the critical part of the
thermal expansion parallels the $T$-derivative of the magnetization
\begin{align}
\alpha_{\rm cr} = \Omega_m \frac{\partial M_{\rm cr}}{\partial T},
\end{align}
where $M_{\rm cr} = - \partial \mathcal{F}_{\rm cr}/\partial H$. In addition, the critical parts of the susceptibility, magnetostriction and compressibility are expected to be proportional
\begin{align} \label{Prop1}
\chi_{\rm cr} = \frac{1}{\Omega_m} \lambda_{\rm cr} = \frac{1}{\Omega_m^2} \kappa_{\rm cr}.
\end{align}
%
Such proportionalities have been observed in CeRu$_2$Si$_2$\cite{Matsuhira99,Bruls90} and also in Sr$_3$Ru$_2$O$_7$.\cite{Grigera04} For a QCEP the susceptibility diverges by definition (\ref{defQCEP}), which in turn implies a divergence in the compressibility $\kappa$. This means that the crystal lattice is destabilized by strong metamagnetic fluctuations, and, as a consequence, the QCEP is likely to be preempted by a structural transition.\cite{Garst1Order,Levanyuk70} We interpret the enormous crystal softening of up to 50\% observed\cite{Bruls90,Flouquet95} in CeRu$_2$Si$_2$ as a precursor of such a structural instability driven by metamagnetic fluctuations.

A strong increase of $\chi_{\rm cr}(T)$ with decreasing $T$, Eq.~(\ref{defQCEP}), also has implications for the $H$-dependence of the specific heat coefficient, $\gamma_{\rm cr}(H)$. Using the higher-order Maxwell relation
\begin{align}  \label{Curvatures}
\frac{\partial^2 \gamma_{\rm cr}}{\partial H^2} = \frac{\partial^2 \chi_{\rm cr}}{\partial T^2},
\end{align}
it directly follows from a positive curvature of $\chi_{\rm cr}(T)$
that $\gamma_{\rm cr}(H)$ exhibits a characteristic minimum at
$H_m$. This implies that the function $\gamma_{\rm cr}(H)$ must
first increase with increasing distance from $H_m$. It is clear that
this increase of $\gamma_{\rm cr}(H)$ cannot continue indefinitely
so that the minimum at the critical field $H_m$ is likely to be
framed by two side peaks at finite $h$. Such a characteristic double
peak structure is observed in CeRu$_2$Si$_2$\cite{Aoki98} and, for
thermodynamic consistency, we predict it to occur in
Sr$_3$Ru$_2$O$_7$ as well.\cite{Rost09} The saturation of $\chi$ in
CeRu$_2$Si$_2$ at a temperature $T^*$ is, according to
Eq.~(\ref{Curvatures}), accompanied by a crossover from a minimum to a maximum in $\gamma(H)$ close to $H_m$ in agreement with
experimental observations.\cite{Aoki98} In addition, from the side peaks of the specific heat coefficient $\gamma_{\rm cr}(H)$ we can infer with the help of the relation
\begin{align} \label{peakgammaalpha}
\frac{\partial \gamma_{\rm cr}}{\partial H} = \frac{\partial^2 M_{\rm cr}}{\partial T^2}
= \frac{1}{\Omega_m} \frac{\partial \alpha_{\rm cr}}{\partial T},
\end{align}
a peak in the temperature dependence of the thermal expansion $\alpha_{\rm cr}(T)$ at the same finite $h$.



Finally, we can exploit the Ising symmetry of the QCEP that demands that the free energy is an even function of the scaling field $h$
%
\begin{align} \label{Ising}
\mathcal{F}_{\rm cr}(h,T,r) = \mathcal{F}_{\rm cr}(-h,T,r).
\end{align}
This Ising symmetry implies, in particular, a sign change of the thermal expansion, $\alpha_{\rm cr}(h) = - \alpha_{\rm cr}(-h)$, at $h=0$. The position of the sign change in $\alpha$ is a good indicator for the strength of the sub-leading temperature dependence of the magnetic scaling field $h(T)$. If this temperature dependence can be neglected, the sign change of $\alpha$ is located exactly at $H=H_m$ and the Ising symmetry is explicit in the phase diagram. A finite temperature dependence will shift the sign change away from $H_m$. Above, we argued that this $T$-dependence will be superlinear due to the Clausius-Clapeyron relation.
%
%
Characteristic sign changes in $\alpha(H)$ and peaks in $\alpha(T)$, see Eq.~(\ref{peakgammaalpha}), have been observed in the metamagnetic materials CeRu$_2$Si$_2$,\cite{Lacerda89b} Sr$_3$Ru$_2$O$_7$\cite{Gegenwart06} and Ca$_{1.8}$Sr$_{0.2}$RuO$_4$.\cite{Baier07} It has been shown\cite{Garst05} that such sign changes of the thermal expansion are generic for quantum criticality and reflect the accumulation of entropy.

In the presence of a QCEP, $r=0$, it follows from general arguments\cite{Zhu03} that the Gr\"uneisen parameter defined as the ratio of thermal expansion and specific heat, $\Gamma_{\rm cr} = \alpha_{\rm cr}/(\gamma_{\rm cr} T)$, saturates in the low temperature limit to a value given by
\begin{align} \label{GammaDiv}
\left.\Gamma_{\rm cr}\right|_{\rm QCEP} \overset{T\to 0}{\longrightarrow} \Omega_m \frac{G}{H - H_m}
\end{align}
where the prefactor $G$ is a combination of critical
exponents.\cite{Zhu03} The sign change in $\Gamma$ at $H_m$ reflects
again the one mentioned above in the thermal expansion. Note that if
the QCEP is only approximately realized, $r>0$, as in CeRu$_2$Si$_2$
the divergence Eq.~(\ref{GammaDiv}), will be cutoff sufficiently
close to $H_m$ and the sign change of $\Gamma$ will not go through
infinity but through zero instead.\cite{Garst1Order} The quantum
critical enhancement of $\Gamma$, Eq.~(\ref{GammaDiv}), offers a
convenient explanation for the anomalous large values of the
Gr\"uneisen parameter observed in
CeRu$_2$Si$_2$.\cite{Lacerda89,Flouquet05}

At some finite distance to the QCEP, the critical free energy
(\ref{Scaling}) will be an analytic function of $h$ in the limit $h
\to 0$, and the limiting behavior
%
%
\begin{align} \label{Exp}
\left.\mathcal{F}_{\rm cr} \right|_{T>0} \approx f_{\rm cr}(T) -
\frac{1}{2} \chi_{\rm max}(T) h^2 + \mathcal{O}(h^4),
\end{align}
is expected.
For positive $r>0$, this expansion should also apply at $T=0$. Note
that the expansion starts quadratically in $h$ due to the Ising
symmetry (\ref{Ising}). The next-order correction of order
$\mathcal{O}(h^4)$ is positive, so that $\chi_{\rm max}(T)$
corresponds to the maximum value of $\chi_{\rm cr}(h)$ at fixed
temperature $T$, see Eq.~(\ref{defQCEP}). The expression (\ref{Exp})
implies some interesting relations between thermodynamic quantities.
The susceptibility for vanishing scaling field $h$, $\chi_{\rm
max}(T) = \chi_{\rm cr}(h=0, T)$, does not only determine the
curvature of $\gamma_{\rm cr}(H)$ close to $H_m$, see
Eq.~(\ref{Curvatures}), but also defines the limiting form of the
thermal expansion
\begin{align} \label{alphaQCR}
\left.\alpha_{\rm cr}\right|_{T>0} \approx \Omega_m \chi'_{\rm max}(T) h + \mathcal{O}(h^3)
\end{align}
where $\chi'_{\rm max} = \partial_T \chi_{\rm max}$. The thermal
expansion is expected to depend linearly on $h \approx H-H_m$ with a
prefactor given by the derivative of the critical differential
susceptibility. Note, however, that we neglect in
Eq.~(\ref{alphaQCR}) contributions from the superlinear temperature
dependence of the magnetic scaling field $h(T)$.

\section{Comparison with $\rm\bf CeRu_2Si_2$}
\label{sec:comparison}

In the following, we discuss in more detail the signatures in thermal expansion, magnetostriction and specific heat of CeRu$_2$Si$_2$ close to the metamagnetic field $H_m$.

\begin{figure}
\includegraphics[width=0.5\textwidth]{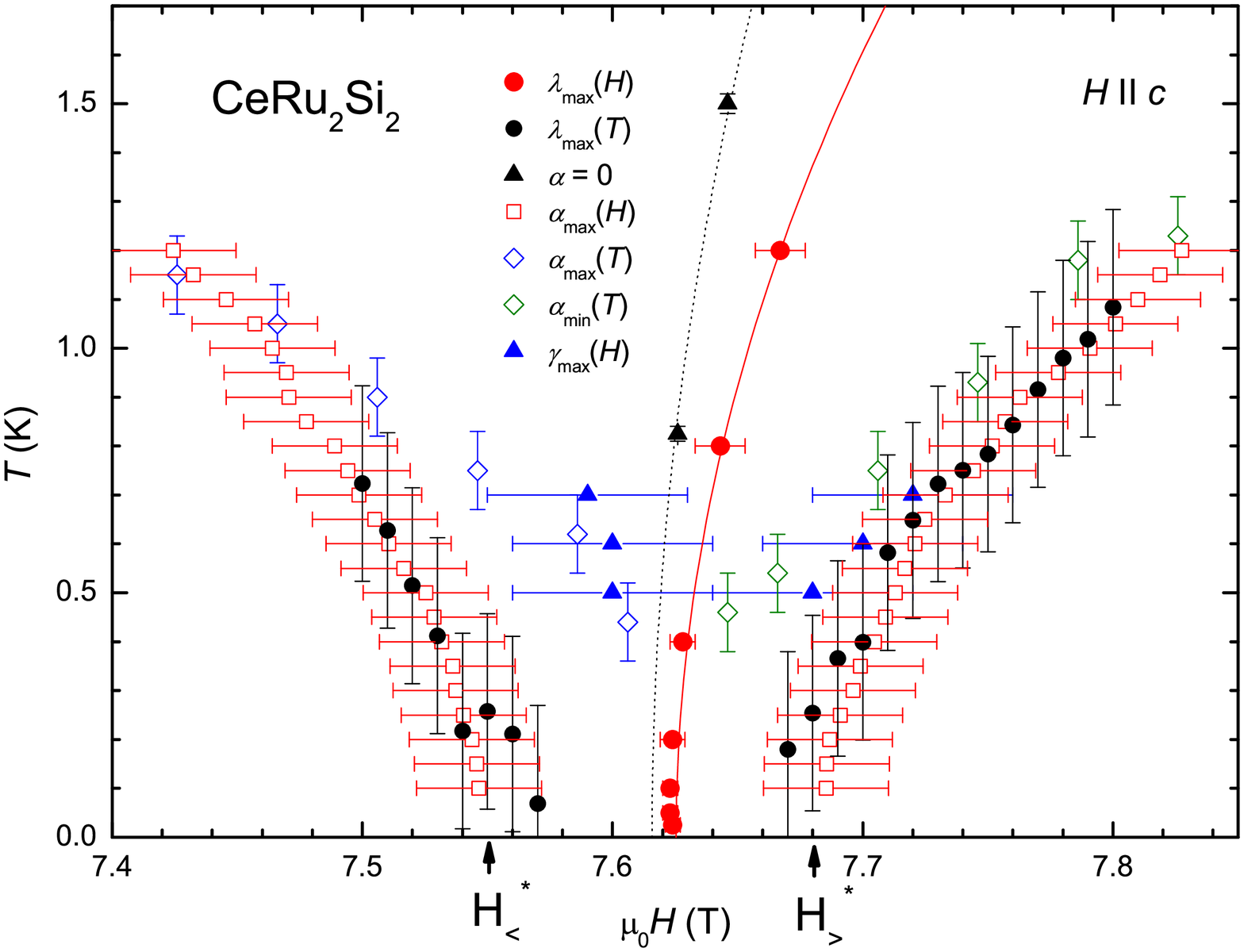}
\caption{Magnetic field-temperature $(H,T)$ plane of CeRu$_2$Si$_2$
close to its metamagnetic field $\mu_{0} H_m = 7.62\,$H with
the positions of extrema in magnetostriction, $\lambda(H)$ and
$\lambda(T)$, thermal expansion, $\alpha(H)$ and $\alpha(T)$,
specific heat $\gamma(H)$, and the positions of vanishing thermal
expansion. Note that the drift of the positions of maxima in
$\lambda(H)$, away from $H_m$ spoils the explicit Ising symmetry.
Critical divergencies in thermodynamics are cutoff upon entering the
pocket enclosed by the positions of extrema in $\alpha(H)$ and
$\alpha(T)$ close to $H_m$; its extention can be quantified by the
temperature and field scale, $T^* = 0.5\,$K and $h^* = (H^*_> -
H^*_<)/2 = 0.07\,$T$/\mu_{0}$, respectively, which are measures of
the distance to the QCEP in parameter space, see text. }
\label{fig:PD}
\end{figure}

The phase diagram close to $H_m$ in Fig.~\ref{fig:PD} summarizes the
positions of maxima and minima in magnetostriction, $\lambda(T)$ and
$\lambda(H)$, thermal expansion, $\alpha(T)$ and $\alpha(H)$, and
specific heat, $\gamma(H)$ and also shows where the thermal
expansion becomes zero. For $T\to 0$, the positions of the maxima in
magnetostriction $\lambda(H)$ approach and, thus, identify the
critical field $\mu_{0}H_m \approx 7.62\,$T. At finite temperatures,
the positions of these maxima and, similarly, the positions of
vanishing thermal expansion deviate from $H_m$ by a distance
proportional to $T^2$, shown by the solid and dotted line,
respectively. We interpret this deviation as arising from a
$T$-dependence of the magnetic scaling field $h(T)$ that spoils the
explicit Ising symmetry in the phase diagram.

\begin{figure}
\includegraphics[width=0.5\textwidth]{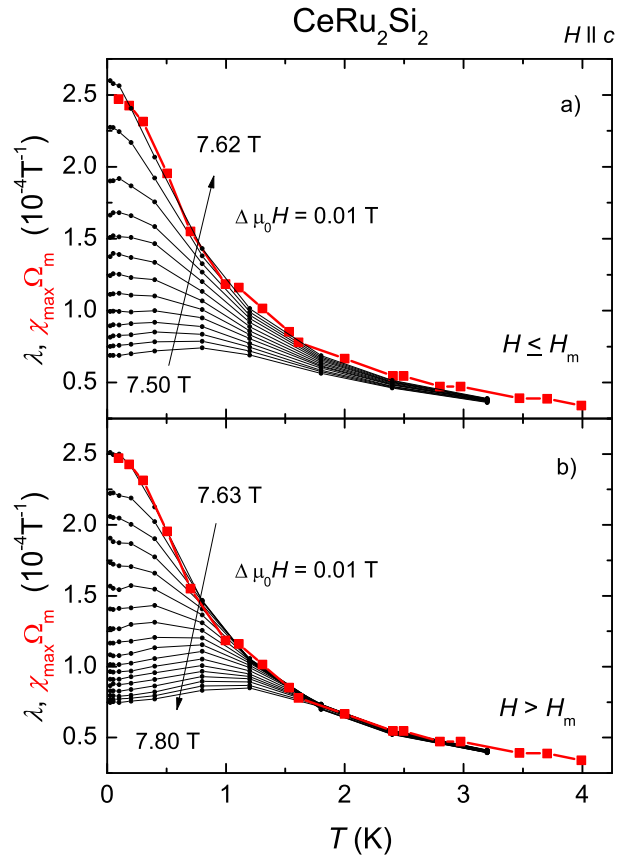}
\caption{Magnetostriction $\lambda(T)$ for different magnetic fields
close to $\mu_{0}H_m = 7.62\,$T; panels (a) and (b) show
$\lambda(T)$ for fields $H \leq H_m$ and $H > H_m$, respectively, in
steps of $0.01\,$T. For comparison, $T$-dependence of maxima
$\chi_{\rm max}(T)$ in the differential susceptibility $\chi(H)$
from Ref.~\onlinecite{Flouquet95} are shown identifying a
proportionality factor $\Omega_m$ = 1.5$\,\pm\,$0.1 T\,kbar$^{-1}$,
see Eq.~(\ref{Prop1}). } \label{fig:lambda}
\end{figure}

{\em magnetostriction --} Magnetostriction as a function of field,
$\lambda(H)$, has been already presented in
Ref.~\onlinecite{Lacerda89} and \onlinecite{Matsuhira99}. In
Fig.~\ref{fig:lambda}, we focus, alternatively, on the temperature
dependence of our magnetostriction data for different fields close
to $H_m$. Upon decreasing temperature, $\lambda(T)$ first increases.
Away from the critical field, $\lambda(T)$ reaches a maximum and
then decreases again until it saturates at a constant value in the
limit $T\to 0$. For fields close to $H_m$, this maximum however
disappears and $\lambda(T)$ increases monotonously with decreasing
$T$. The magnetostriction close to the critical field has the
largest absolute values identified by the envelope, $\lambda_{\rm
max}(T)$, of the set of $\lambda(T)$-curves in
Fig.~\ref{fig:lambda}. In the following arguments, the
proportionality of magnetostriction and differential susceptibility
$\chi$ will be of importance; this correspondence was impressively
demonstrated in Ref.~\onlinecite{Matsuhira99}. Here, we show this
correspondence again by comparing the envelope of
$\lambda(T)$-curves with the temperature dependence of the maxima in
$\chi(H)$-data taken from Ref.~\onlinecite{Flouquet95} (red squares
in Fig.~\ref{fig:lambda}). The proportionality factor determines
$\Omega_m$ as defined in Eq.~(\ref{OmegaDef}). We obtain the value
$\Omega_m$ = 1.5$\,\pm\,$0.1 T\,kbar$^{-1}$, which is slightly
smaller than the value of 2.0 T\,kbar$^{-1}$ estimated by pressure
experiments. \cite{Aoki01} The magnetostriction at the critical
field saturates below the characteristic temperature $T^* = 0.5\,$K,
which we identify as a crossover from critical to non-critical
behavior associated with a finite distance to the QCEP in parameter
space. As we will see below, the inflection point of $\lambda(T)$
located at $T^*$ will also determine crossover signatures in other
thermodynamic quantities.

The positions of the characteristic maxima in $\lambda(T)$ are shown
in the $(H,T)$-plane of Fig.~\ref{fig:PD}. Interestingly, the maxima
in $\lambda(T)$ and $\chi(T)$ imply, according to the relation
\begin{align}
\frac{\partial \lambda_{\rm cr}}{\partial T} = \Omega_m
\frac{\partial \chi_{\rm cr}}{\partial T} = \frac{\partial
\alpha_{\rm cr}}{\partial H},
\end{align}
an extremum in the field dependence of the thermal expansion,
$\alpha_{\rm cr}(H)$, at the same positions in the phase diagram,
which we will confirm below. Note that characteristic maxima in the
temperature dependence of $\chi(T)$, which are, according to
Eq.~(\ref{Prop1}), equivalent to the maxima in $\lambda(T)$ of
Fig.~\ref{fig:lambda}, have been also observed in the metamagnetic
material Sr$_3$Ru$_2$O$_7$.\cite{Ikeda00}

\begin{figure}
\includegraphics[width=0.5\textwidth]{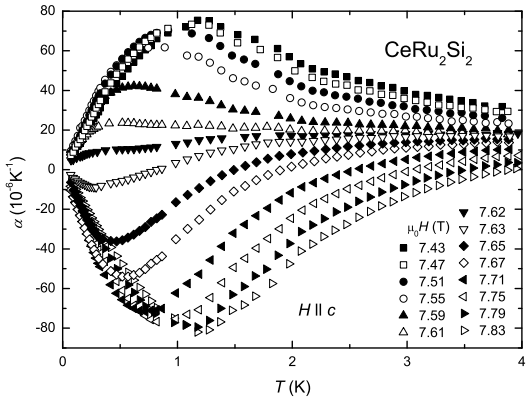}
\caption{Thermal expansion as a function of temperature for various
magnetic fields close to $\mu_{0}H_m = 7.62\,$T. The curves are
approximately mirror-symmetric due to the emergent Ising symmetry of
the QCEP.} \label{fig:alpha}
\end{figure}

{\em thermal expansion --} Thermal expansion data of CeRu$_2$Si$_2$
has been presented in Refs.~\onlinecite{Lacerda89b} and
\onlinecite{Paulsen90}. Already in zero field, $H=0$, $\alpha(T)$
exhibits a peak
that shifts to lower temperatures and sharpens with increasing $H$. At the critical field $H_m$, thermal expansion
changes sign, and the negative peak in $\alpha(T)$ broadens
and shifts to higher temperatures for increasing fields
$H>H_m$. This behavior is reminiscent of quantum critical
metamagnetism as discussed in some detail in
Ref.~\onlinecite{Gegenwart06}. Similar behavior of $\alpha(T)$ is
observed near the metamagnetic field of
Sr$_3$Ru$_2$O$_7$\cite{Gegenwart06} and
Ca$_{1.8}$Sr$_{0.2}$RuO$_4$.\cite{Baier07} A closer inspection,
however, reveals that the behavior very close to the critical field
differs qualitatively from the theoretical expectations for a QCEP,
see Fig.~\ref{fig:alpha}. When the peak position has reached a
temperature of the order of $T^*=0.5\,$K, it does not shift further
towards lower temperatures upon increasing $H$, but its height
instead decreases to zero and reemerges with opposite sign for
fields $H>H_m$. The thermal expansion curve $\alpha(T)$ then almost
recovers its shape but with opposite sign. We interpret this
qualitative change as a crossover from critical to non-critical
behavior associated with the temperature scale $T^*$. For
temperatures $T<T^*$, the thermal expansion has temperature
dependence $\alpha \propto T$ characteristic for a Fermi-liquid.

\begin{figure}
\includegraphics[width=0.5\textwidth]{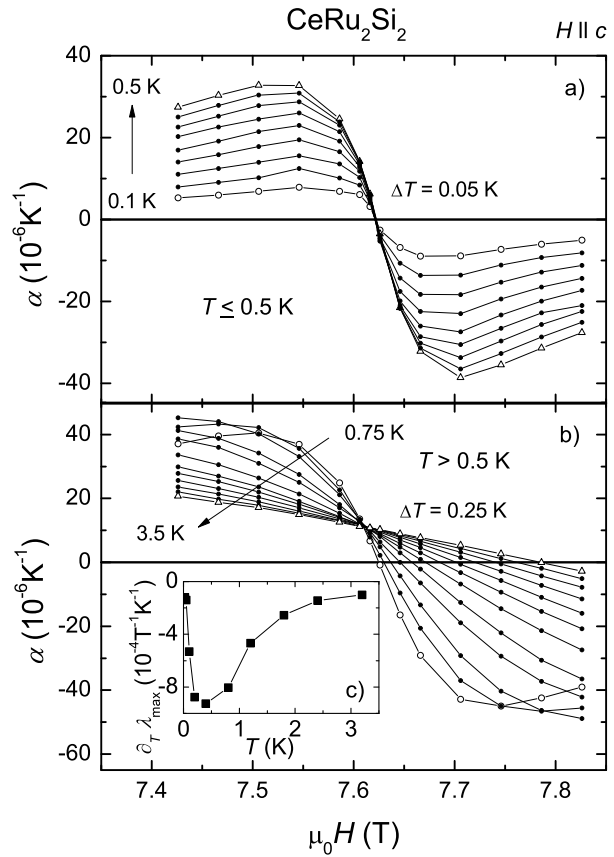}
\caption{Thermal expansion as a function of field $H$ for various
temperatures obtained by extrapolation from the data set in
Fig.~\ref{fig:alpha}. Panel (a) and (b) show temperatures $T\leq T^*$
and $T>T^*$, respectively, with $T^*=0.5\,$K and steps $\Delta T =
0.05\,$K and $\Delta T = 0.25\,$K, respectively. Their slope close
to the critical field is given by $\partial_H \alpha|_{H=H_m}
\approx
\partial_T \lambda_{\rm max}$ whose $T$-dependence is shown in the
inset (c). Its extremal value for the slope, min$\{\partial_T
\lambda_{\rm max}\} \approx -0.001\,$(T K)$^{-1}$, at $T^*$
determines the bundling slope of the $\alpha(H)$-curves close to
$H_m$ in panel (a).} \label{fig:alphaComp}
\end{figure}

The dense data set in Fig.~\ref{fig:alpha} allows to discuss
the magnetic field dependence of the thermal expansion at a given
temperature, $\alpha(H)$. The extrapolated curves are shown in
Fig.~\ref{fig:alphaComp}(a) and (b) for temperatures $T\leq T^*$ and
$T>T^*$, respectively. For low temperatures, $T\leq T^*$, $\alpha(H)$
has a point reflection symmetry located at $H=H_m$ and $\alpha = 0$
which is characteristic for an emergent Ising symmetry close to a
critical endpoint. The absolute value of thermal expansion
$|\alpha(H)|$ first increases upon approaching the critical field
$H_m$ but after reaching an extremum it decreases and vanishes at
$H_m$. The positions of these extrema in $\alpha(H)$ are shown in
Fig.~\ref{fig:PD}. As argued above, these positions in the phase
diagram coincide with the positions of corresponding maxima in
$\lambda(T)$. In the low temperature limit they extrapolate to the
fields $\mu_{0}H^*_< = 7.55\,$T and $\mu_{0}H^*_> = 7.68\,$T; their
distance identifies a finite field scale $h^* = (H^*_> -H^*_<)/2 =
0.07\,$T$\mu_{0}^{-1}$ attributed to the vanishing of the extrema in
$\lambda(T)$ and $\alpha(H)$. This finite field scale $\mu_{0}h^* =
0.07\,$T indicates that the increase of $|\alpha(H)|$ while
approaching $H_m$ always gives way to a decrease even at lowest
temperatures. Hence,
it signifies a crossover
from critical to non-critical behavior at $T=0$ and is thus a
magnetic field analog of the temperature scale $T^*$.

Furthermore, note that the curves $\alpha(H)$ in
Fig.~\ref{fig:alphaComp}(a) tend to bundle near $H_m$ to a line with
constant slope. Generally, the slope of $\alpha(H)$ can be
identified with the derivative of magnetostriction $\partial_T
\lambda$ or, equivalently, of the susceptibility $\partial_T \chi$.
The numerical derivative of $\lambda_{\rm max}(T)$, the envelope in
Fig.~\ref{fig:lambda}, is shown in Fig.~\ref{fig:alphaComp}(c). It is
its minimum value, min$\{\partial_T \lambda_{\rm max}\} =
-0.001\,$(T K)$^{-1}$, at $T^*$ that identifies the bundling slope
in Fig.~\ref{fig:alphaComp}. Moreover, the two lines in the phase
diagram that identify the extrema of $\alpha(H)$ are also the
boundaries of the region where the analytic expansion of the
critical free energy in the scaling field (\ref{Exp}) holds. Well
within this region the thermal expansion depends linearly on $H-H_m$
as expected from Eq.~(\ref{alphaQCR}).

For larger temperatures, $T>T^*$, the thermal expansion $\alpha(H)$ is shown in Fig.~\ref{fig:alphaComp}(b). Apparently, the field range where $\alpha(H)$ is approximately linear increases, see Eq.~(\ref{alphaQCR}), its slope decaying with increasing temperature in accordance with the behavior of $\partial_T \lambda_{\rm max}$. In contrast to the low-temperature limit, the curves do not intersect at $H=H_m$ and $\alpha=0$ anymore, which we attribute to the incipient $T$-dependence of the magnetic scaling field $h$ that breaks the explicit Ising symmetry.


\begin{figure}
\includegraphics[width=0.5\textwidth]{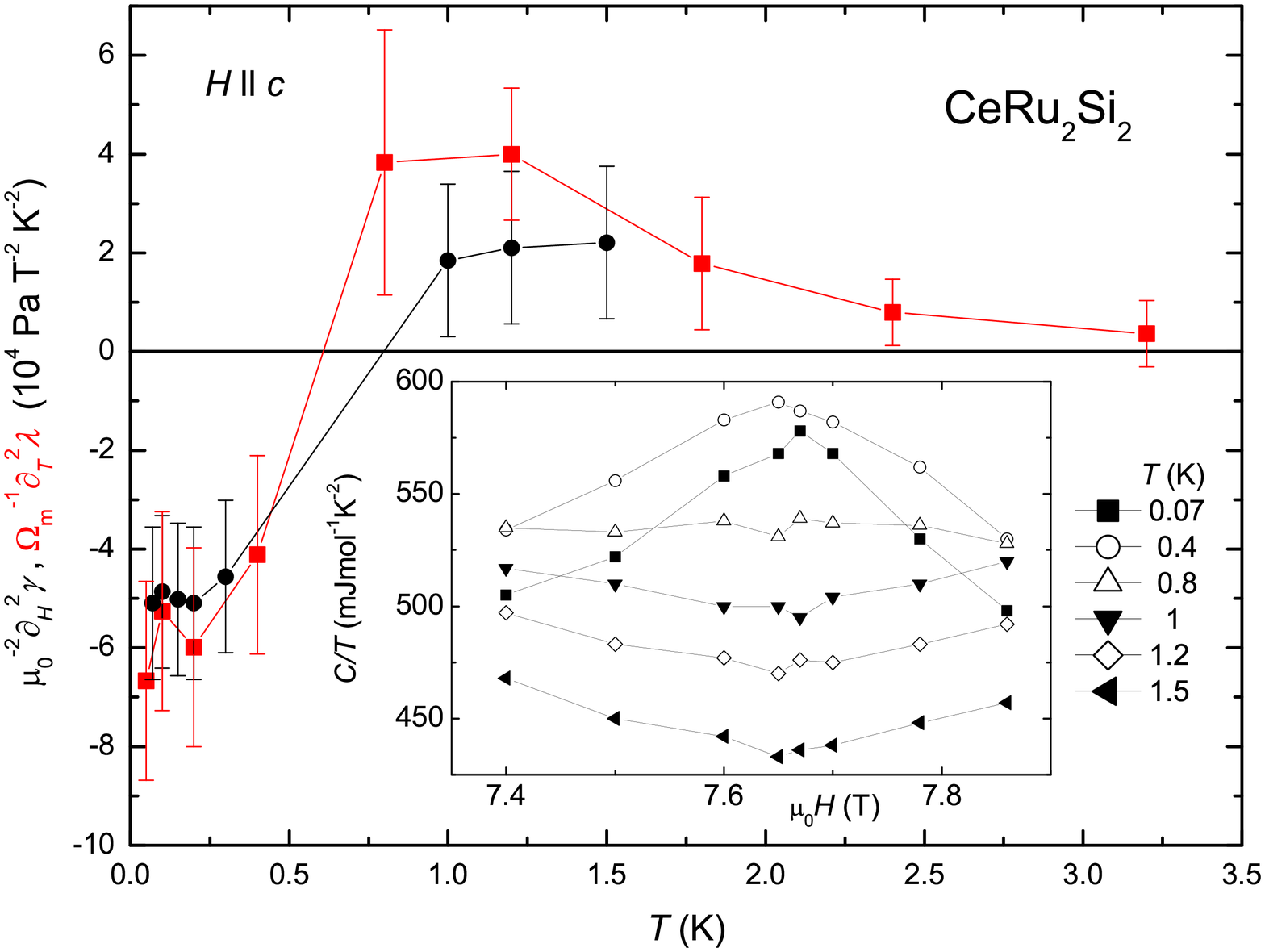}
\caption{The specific heat coefficient $\gamma(H)$ exhibits a
minimum/maximum crossover close to the critical field $H_m$ with
decreasing temperature, see inset. The main panel shows the
curvature $\partial^2_H \gamma(H)$ (black circles) at the extremum
close to $H_m$ as a function of $T$, which is compared to
$\partial^2_T \lambda_{\rm max}/\Omega_m$ (red squares). By virtue
of Eq.~(\ref{Curvatures}), the minimum/maximum crossover of
$\gamma(H)$ is determined by an inflection point in the
susceptibility, $\partial^2_T\chi_{\rm max} = \partial^2_T
\lambda_{\rm max}\Omega_m^{-1} = 0$, at $T^*= 0.5\,$K. }
\label{fig:gammaComp}
\end{figure}

{\em specific heat --} Specific heat data have been reported in
Ref.~\onlinecite{Aoki98}. The specific heat coefficient $\gamma$ is
enhanced close to the critical field by almost a factor of two
compared to its zero field value. However, one observes a sharp
single peak close to $H_m$ only at lowest temperatures, $T<T^*$. At
elevated temperatures, $T>T^*$, a double peak structure in
$\gamma(H)$ with a minimum at $H_m$ is found. In the following, we
focus on the behavior of $\gamma$ close to the critical field (see
inset of Fig.~\ref{fig:gammaComp}) and, in particular, analyze the
minimum/maximum crossover at $T^*$. The fitted curvatures of
$\gamma(H)$ at the extremum close to $H_m$ are shown in
Fig.~\ref{fig:gammaComp} for each available temperature. The result
is compared to the numerical second order derivative of
$\lambda_{\rm max}(T)$ or, equivalently, $\chi_{\rm max}(T)$ of
Fig.~\ref{fig:lambda}. As anticipated from Eq.~(\ref{Curvatures}),
the two quantities agree within the error bars. In particular, the
minimum/maximum crossover is identified with the {\em inflection
point} of $\chi_{\rm max}(T)$ at $T\approx T^*$ where $\partial^2_T
\chi_{\rm max}(T) = 0$. In the phase diagram Fig.~\ref{fig:PD}, we
also show the positions of the side peaks in $\gamma(H)$ (not shown in Fig.~\ref{fig:gammaComp}) that
coincide within the error bars with the positions of the maxima in
$\alpha(T)$ as expected from Eq.~(\ref{peakgammaalpha}).

\section{Summary}
\label{sec:summary}

We presented a comprehensive discussion of the universal
thermodynamic signatures, which emerge close to a metamagnetic
quantum critical endpoint. We argued that (i) the diverging
differential susceptibility together with (ii) the Ising symmetry of
the QCEP account for the following characteristics of critical
metamagnetism: (1) a proportionality between susceptibility,
magnetostriction and compressibility in the presence of a
magnetoelastic coupling and, as a result, (2) a pronounced crystal
softening, (3) an enhanced Gr\"uneisen parameter, (4) a sign change
of the thermal expansion at the critical field, and (5) a minimum in
the specific heat coefficient $\gamma(H)$. The latter minimum
directly follows from the  Maxwell relation $\partial_T^2 \chi =
\partial^2_H \gamma$ and the positive curvature of the
susceptibility. A minimum in $\gamma(H)$, in turn, implies two side
peaks, and we showed that their positions in the $(H,T)$ plane
coincide with the extrema in the thermal expansion $\alpha(T)$ as a
function of temperature. The Ising symmetry of the endpoint ensures
that the sign change of the thermal expansion $\alpha(H)$ due to
entropy accumulation\cite{Garst05} occurs close to the critical
field $H_m$.

As an example, we discussed the metamagnetic compound
CeRu$_2$Si$_2$, which shows pronounced metamagnetic signatures that
saturate, however, at very low temperatures, $T^* = 0.5\,$K, and
close to the critical field, $|H-H_m| < h^*= 0.07\,$T$\mu_{0}^{-1}$.
We argued that outside this regime in the phase diagram behavior for
a metamagnetic QCEP is expected. We presented new high-precision
data close to the critical field, that allowed us, in particular, to
analyze the crossover from critical to non-critical behavior
associated with the temperature and field scale, $T^*$ and $h^*$,
respectively. We demonstrated that the onset of saturation in the
differential susceptibility $\chi(T)|_{H=H_m}$ at $T^*$ leads to an
inflection point that accounts for the minimum-to-maximum crossover
in the specific heat coefficient $\gamma(H)$ at $H_m$. Furthermore,
we demonstrated that the derivative $\partial_T\chi(T)|_{H=H_m}$
determines the temperature dependence of the thermal expansion close
to the critical field. The inflection point at $T^*$ is reflected by
an extremum in $\alpha(T)$ at the same temperature as the critical
field is approached. We also demonstrated that the magnetostriction
$\lambda(T)$ and, equivalently, the differential susceptibility
$\chi(T)$ has a maximum as a function of temperature only for fields
$|H-H_m|>h^*$ allowing us to identify the field scale $\mu_{0}h^* =
0.07\,$T. Such characteristic maxima have been observed also in the
metamagnetic compound Sr$_3$Ru$_2$O$_7$.\cite{Ikeda00} We pointed
out that these maxima coincide with thermodynamically equivalent
extrema in the thermal expansion $\alpha(H)$.

The focus of the present work are the qualitative thermodynamic
signatures close to quantum critical metamagnetism, irrespective of
the microscopic details of the material and the precise model
describing the critical dynamics of the order-parameter
fluctuations. We checked, however, that all signatures discussed
here for CeRu$_2$Si$_2$ are reproduced qualitatively within the
QCEP-model introduced in Ref.~\onlinecite{Millis02}, which will be
the subject of a separate publication.

The universal metamagnetic signatures have been here illustrated on the heavy-fermion compound CeRu$_2$Si$_2$. We hope that this work will motivate further experimental investigations to identify additional materials that are close to a metamagnetic QCEP.
 

\acknowledgments We acknowledge P. Haen for providing the high
quality single crystal and J. A. Mydosh for fruitful discussions
about \crs. Discussion with T. Lorentz on metamagnetism in
Ca$_{1.8}$Sr$_{0.2}$RuO$_4$ is gratefully acknowledged as well. The
work was funded by the DFG through FOR 960 and SFB 608 (M.G.) and by
the Max-Planck-Society via project\# M.FE.A.CHPHSM (F.W.).

\appendix

\end{document}